\documentclass[12pt]{article}

\pdfoutput=1
\DeclareFontFamily{T1}{calligra}{}
\DeclareFontShape{T1}{calligra}{m}{n}{<->s*[1.44]callig15}{}
\DeclareMathAlphabet\mathcalligra   {T1}{calligra} {m} {n}
\DeclareMathAlphabet\mathzapf       {T1}{pzc} {mb} {it}
\DeclareMathAlphabet\mathchorus     {T1}{qzc} {m} {n}
\DeclareMathAlphabet\mathrsfso      {U}{rsfso}{m}{n}
\DeclareMathAlphabet\mathfrcal      {T1}{frcursive}{m}{it}
\DeclareFontFamily{T1}{frcursive}{}
\DeclareFontShape{T1}{frcursive}{m}{n}{<->s*[1.44]callig15}{}
\DeclareMathAlphabet\mathfrcal      {T1}{frcursive}{m}{it}

\usepackage{amsmath}
\usepackage{amssymb}
\usepackage{graphicx}
\usepackage[dvipsnames]{xcolor}
\usepackage{soul}
\numberwithin{equation}{section}
\usepackage{array}
\usepackage{mathtools}
\usepackage{dsfont}
\usepackage{mathrsfs}

\usepackage{BOONDOX-uprscr}

\usepackage{tikz}\usetikzlibrary{matrix,fit}
\usepackage{tikz-cd} 
\usepackage{varwidth}
\usepackage{enumerate}
\usepackage{appendix}
\usepackage{xfrac}
\usepackage{nicefrac}
\usepackage{mathtools,slashed}

\usepackage[margin=1in]{geometry}
\usepackage{nicematrix}

\usepackage[
    backend=bibtex,
    style=alphabetic,
maxbibnames=99,
giveninits=true,
minalphanames=1,
maxalphanames=3,url=false
  ]{biblatex}

\bibliography{SUSYGrassmannian}

\usepackage{mathabx}
\usepackage{empheq}

\usepackage{setspace}

\usepackage{slashed}
\usepackage{upgreek}
\usepackage{appendix}

\usepackage{tabstackengine}

\usepackage{wrapfig}
\usepackage[abs]{overpic}

\usepackage{float}

\fixTABwidth{T}

\newdimen\mytextwidth
\newcommand\rem[2][cyan!40!green]{\noindent\nobreak\hfil\penalty1000\hfilneg
\mytextwidth=\linewidth\advance\mytextwidth by 2mm
\begin{tikzpicture}[baseline=-\the\dimexpr\fontdimen22\textfont2\relax]\node[outer sep=0pt,draw=black,fill=#1,fill opacity=1,text opacity=1,rectangle,rounded corners]{\begin{varwidth}{\mytextwidth}\textcolor{white}{#2}\end{varwidth}};
\end{tikzpicture}\allowbreak
}

\newcommand\whiterem[2][white!]{\noindent\nobreak\hfil\penalty1000\hfilneg
\mytextwidth=\linewidth\advance\mytextwidth by 2mm
\begin{tikzpicture}[baseline=-\the\dimexpr\fontdimen22\textfont2\relax]\node[outer sep=0pt,draw=black,fill=#1,fill opacity=1,text opacity=1,rectangle,rounded corners,line width=1.5pt]{\begin{varwidth}{\mytextwidth}\textcolor{black}{#2}\end{varwidth}};
\end{tikzpicture}\allowbreak
}

\makeatletter
\newsavebox{\@brx}
\newcommand{\llangle}[1][]{\savebox{\@brx}{\(\m@th{#1\langle}\)}%
  \mathopen{\copy\@brx\kern-0.5\wd\@brx\usebox{\@brx}}}
\newcommand{\rrangle}[1][]{\savebox{\@brx}{\(\m@th{#1\rangle}\)}%
  \mathclose{\copy\@brx\kern-0.5\wd\@brx\usebox{\@brx}}}
\makeatother

\renewcommand{\tilde}{\widetilde}

\newcommand{\bea}{\begin{equation}}
\newcommand{\eea}{\end{equation}}
\newcommand{\bear}{\begin{eqnarray}}
\newcommand{\eear}{\end{eqnarray}}
\newcommand{\bearr}{\begin{eqnarray*}}
\newcommand{\eearr}{\end{eqnarray*}}

\newcommand{\dif}{\mathrm{d}}

\usepackage{ytableau}
\ytableausetup{centertableaux}

\usepackage{tikz}
\usepackage{xparse}
\NewDocumentCommand{\xrightarrows}{ O{}O{} }{%
\mathrel{%
\vcenter{\hbox{%
\begin{tikzpicture}
  \node[minimum width=1cm,minimum height=1ex,anchor=south,align=center] (a){\text{\vphantom{hg}#1}\\[0.5ex] \vphantom{hg}#2};
  \draw[<-] ([yshift=0.35ex]a.west) -- ([yshift=0.35ex]a.east);
  \draw[->] ([yshift=-0.35ex]a.west) -- ([yshift=-0.35ex]a.east);
\end{tikzpicture}
}}%
}%
}

\renewbibmacro{in:}{}

\usepackage{mdframed}

\newbibmacro{string+doi}[1]{
  \iffieldundef{doi}{#1}{\href{http://dx.doi.org/\thefield{doi}}{#1}}}
\DeclareFieldFormat{title}{\usebibmacro{string+doi}{\mkbibemph{#1}}}
\DeclareFieldFormat[article]{title}{\usebibmacro{string+doi}{\mkbibquote{#1}}}

\newmdenv[
  topline=false,
  bottomline=false,
  rightline=false,
  linewidth=2pt,
  skipabove=\topsep,
  skipbelow=\topsep
]{siderules}

\newmdenv[
  topline=false,
  bottomline=false,
  linewidth=2pt,
  skipabove=\topsep,
  skipbelow=\topsep
]{siderulesright}

\makeatletter
\renewcommand{\@seccntformat}[1]{\csname the#1\endcsname.\quad}
\makeatother

\usepackage{setspace}
\usepackage{xpatch}

\makeatletter
\renewcommand{\@chap@pppage}{
  \clear@ppage
  \thispagestyle{plain}
  \if@twocolumn\onecolumn\@tempswatrue\else\@tempswafalse\fi
  \null\vfil
  \markboth{}{}
  {\centering
   \interlinepenalty \@M
   \normalfont
   \MakeUppercase \appendixpagename\par}
  \if@dotoc@pp
    \addappheadtotoc
  \fi
  \vfil\newpage
  \if@twoside
    \if@openright
      \null
      \thispagestyle{empty}
      \newpage
    \fi
  \fi
  \if@tempswa
    \twocolumn
  \fi
}
\makeatother

\definecolor{navycol}{RGB}{100,150,160}
   \definecolor{pinkcol}{RGB}{242,55,55}
   \definecolor{greencol}{RGB}{50,205,50}

   \definecolor{bluecol}{RGB}{30,144,255}

\usepackage{titlesec}

\titleformat*{\section}{\large\bfseries}
\titleformat*{\subsection}{\normalsize\bfseries}
\titleformat*{\subsubsection}{\normalsize\bfseries}
\titleformat*{\paragraph}{\large\bfseries}
\titleformat*{\subparagraph}{\large\bfseries}
\titlespacing{\author}{-5pt}{-5pt}{-5pt}[-5pt]

\makeatletter
\renewcommand\subsubsection{\@startsection{subsubsection}{3}{\z@}
                                     {-3.25ex\@plus -1ex \@minus -.2ex}
                                     {-1.5ex \@plus -.2ex}
                                     {\normalfont\normalsize\bfseries}}
\renewcommand\subsection{\@startsection{subsection}{3}{\z@}
                                     {-3.25ex\@plus -1ex \@minus -.2ex}
                                     {-1.5ex \@plus -.2ex}
                                     {\normalfont\normalsize\bfseries}}                                     
\makeatother

\DeclareFontFamily{U}{solomos}{}
\DeclareErrorFont{U}{solomos}{m}{n}{10}
\DeclareFontShape{U}{solomos}{m}{n}{
  <-> s*[1.1]  gsolomos8r
}{}

   \usepackage{stmaryrd}

\usepackage{tikz}
\usetikzlibrary{arrows.meta}

\usepackage{indentfirst}

\usepackage{tocloft}
\let \savenumberline \numberline
\def \numberline#1{\savenumberline{#1.}}

\usepackage{etoolbox}
\patchcmd{\tableofcontents}{\@starttoc}{\vspace{-0.3cm}\@starttoc}{}{}

\usepackage{accents}

\newcommand\smallthickbar[1]{\accentset{\rule{.5em}{.7pt}}{#1}}
\usepackage{hyperref}
\hypersetup{
colorlinks=true,
linkcolor=MidnightBlue,
citecolor=violet,
filecolor=purple,
urlcolor=cyan,
breaklinks=true
}

\newcounter{Chapcounter}

\newcommand{\chapter}[1] 
{ {\centering          
  \addtocounter{Chapcounter}{1} \Large \underline{\sffamily \texorpdfstring{\textbf{  Chapter \theChapcounter: ~#1}}{Lg}} }   
  \addcontentsline{toc}{section}{ \color{MidnightBlue} \texorpdfstring{Chapter ~}{Lg}\theChapcounter.\texorpdfstring{~~}{Lg} #1 }    
}

\begin{document}

\title{\vspace{-1.0cm} \textbf{Point Particles as Spin Chains}
\vspace{0.5cm}
}

\author{Viacheslav Krivorol$^{\,\triangle,\,\lambda,\,}$\footnote{Email: v.a.krivorol@gmail.com}
\\  \vspace{0cm}  \\
{\small $\triangle)$ \emph{Institute for Theoretical and Mathematical Physics,}} \\{\small \emph{Lomonosov Moscow State University, 119991 Moscow, Russia}}\\
{\small $\lambda)$ \emph{HSE University, 6 Usacheva str., 119048 Moscow, Russia}}
}

\date{}

{\let\newpage\relax\maketitle}

\maketitle

\vspace{0cm}
\textbf{Abstract.} 
This work surveys a recently developed approach to the study of free point particles on Riemannian manifolds, based on the Kirillov orbit method, geometric quantization, and the geometry of Lagrangian submanifolds. We discuss that given a Lagrangian submanifold $\mathcal{M}$ embedded in a product of coadjoint orbits and a Hamiltonian $H$ attaining its minimum on this submanifold, such a configuration naturally induces free point particle dynamics on $\mathcal{M}$. The metric governing this dynamics is precisely defined by the quadratic expansion of $H$ around its minimum. Upon quantization, this correspondence establishes a relation between the $L^2(\mathcal{M})$ and a corresponding spin chain Hilbert space as well as a spectral equivalence between Laplace-Beltrami operator on $L^2(\mathcal{M})$ and a spin Hamiltonian. Explicit examples of this construction are presented for particles moving on the complex plane, two-dimensional sphere, flag manifolds, and the hyperbolic plane.

\newpage
\tableofcontents

\newpage

\section{Introduction}
Suppose $\mathcal{M}$ is a Riemannian manifold. The central object of our investigation is quantum free point particles on $\mathcal{M}$. From the mathematical point of view, the corresponding Hamiltonians are associated with Laplace-type operators $-\triangle$ acting, in general, on square-integrable sections of vector bundles $\mathscr{E}$ on $\mathcal{M}$. Such operators include the Laplace-Beltrami operator acting on functions $L^2(\mathcal{M})$, the Bochner (or magnetic) Laplacian acting on sections of vector bundles $L^2(\mathcal{M},\mathscr{E})$, the Laplace-de Rham operator acting on differential forms $\Omega^{\bullet}(\mathcal{M})$, among others. One of the main questions in \textit{spectral geometry} \cite{Weyl1912,Kac,RevModPhys.82.2213} is the study of the spectral problem
\begin{equation}\label{LaplaceSpectralProblemGeneral}
-\triangle\Psi = E\Psi\,.
\end{equation}
However, as practice demonstrates, this spectral problem is as rich as it is challenging. Indeed, it constitutes a second-order elliptic partial differential equation for which no general solution algorithm exists. In practice, this problem can be solved exactly only for manifolds with ``sufficiently high'' symmetry, for references see the introduction in \cite{BykovGaudin}.
The survey of one possible approach to this problem, largely motivated by quantum physics, constitutes the aim of the present work.

Indeed, as previously noted, the spectral problem \eqref{LaplaceSpectralProblemGeneral} has a remarkably productive physical interpretation. A number of Laplace-type operators arise as quantum Hamiltonians\footnote{The ordering ambiguity in quantization can, in fact, introduce an additive scalar curvature contribution to the Hamiltonian \cite{Wagner:2021luc}. We will disregard this term, however, as our focus will be on invariant metrics on homogeneous spaces, where the curvature is constant.} of the form $H = -\triangle$ for so-called one-dimensional \textit{sigma models}. These describe the dynamics of essentially free particles moving on a Riemannian manifold $\mathcal{M}$, possibly coupled to additional geometric structures. The standard action for such theories is of the form
\begin{equation}\label{GeneralSigmaModel}
\mathcal{S} = \frac{1}{2}\int \big(g_{ij}\dot{q}^{\,i}\dot{q}^{\,j}\big)\dif t+\int{\mathcal{A}}+\text{(fermions)}\,,
\end{equation}
where $q^i$ are coordinates on $\mathcal{M}$, $g_{ij}(q)$ are the components of the metric tensor, and $\mathcal{A}$ is a 1-form\footnote{It turns out that there are interesting physical situations where the 1-form $\mathcal{A}$ is not defined globally, yet the action $\mathcal{S}$ remains well-defined \cite{Wu:1975es,Alvarez:1984es}. The Dirac monopole on the sphere is the simplest such example \cite{Dirac:1931kp}.} on $\mathcal{M}$, interpretable as the vector potential of a magnetic field. To obtain a Laplace-type operator acting on differential forms upon quantization, fermions are typically introduced in a supersymmetric manner \cite{SmilgaBook}. In this setting, equation \eqref{LaplaceSpectralProblemGeneral} may be interpreted as the stationary Schrödinger equation for the corresponding quantum sigma model.

So far, this reformulation will help us to find a more productive representation for the spectral problem of Laplace operators. Specifically, this will help us to replace the problem of diagonalizing a Laplace operator on a certain manifold with the problem of finding the eigenvalues of some Hamiltonian for a specific \textit{spin chain}. 
In this paper, we refer to quantum spin chains as quantum mechanical systems whose Hilbert spaces are formed by the tensor product of irreducible representations of certain Lie groups. In some cases, such a reformulation can greatly simplify the spectral problem and use new methods typical for spin models, for example, the Bethe Ansatz \cite{Slavnov:2018kfx}. In this formalism, it is often convenient to switch to an oscillator representation \cite{PerelomovBook}, which allows us to reduce the problem to the calculation in terms of ladder operators form the Heisenberg algebra. From a physical point of view, this approach is also beneficial, as it provides a new global quantization scheme for point particles, which means that it does not rely on specific patches of the manifold. This opens up new opportunities for research into areas where such a property might be significant, for example the study of AdS particles \cite{Basile:2023vyg} or even BMS particles \cite{Bekaert:2024uuy,Barnich:2015uva}. Also, as we will see later, this method sometimes allows us to reduce the quantization of non-compact cotangent bundles to the quantization of some compact manifolds, which is interesting in its own right in the context of geometric quantization \cite{Wernli:2023pib,Kirillov2001}. In the complex setting, we can sometimes reformulate problems in \textit{harmonic analysis} \cite{Helgason1981} entirely within the framework of \textit{complex} or even \textit{algebraic geometry} \cite{GriffithsHarris1978}.

So, as previously noted, the aim of this paper is to establish a connection between two, at first glance unrelated, problems:
\begin{equation}
\text{1D Sigma Models}\qquad\leftrightsquigarrow\qquad\text{Spin Chains}\,.
\end{equation}
The underlying approach is to replace the particle's phase space, $\mathrm{T}^\ast\mathcal{M}$ -- which can be challenging to quantize directly -- with a product of simpler manifolds whose geometric quantization is well understood. Specifically, one uses manifolds whose quantization yields a single irreducible unitary representation of a group, corresponding to the simplest, or minimal, possible Hilbert space.
As we know from the theory of geometric quantization, to obtain unitary irreducible representations as the resulting Hilbert spaces, one needs to perform geometric quantization of specific symplectic manifolds known as coadjoint orbits $\mathcal{O}$ endowed with the natural Kirillov-Kostant-Souriau symplectic form\footnote{In practice, we will not use the general definition of a coadjoint orbit and its symplectic structure. Instead, we assume familiarity with the fact that in specific cases they are isomorphic to well-known symplectic manifolds, such as spheres, flag manifolds, hyperbolic planes, etc.} \cite{Kirillov:1999qgl}. In the most common physical case of simply-connected compact Lie groups such as $\mathbf{SU}(n)$, $\mathbf{SO}(n)$, and $\mathbf{Sp}(2n)$, this statement is embodied by the \textit{Borel-Weil-Bott theorem} \cite{Bott1990}. Informally, it asserts that their finite-dimensional irreducible unitary representations can be realized on spaces of holomorphic sections (or, more generally, higher sheaf cohomology) of certain line bundles over generalized flag manifolds. For a broader class of groups, the orbit method remains more of a guiding principle than a complete theory. Nevertheless, analogous constructive results exist for specific group types, such as nilpotent groups \cite{Kirillov1962} (e.g., the Heisenberg group) and certain non-compact groups (e.g., $\mathbf{SL}(2,\mathbb{R})$, see the Section 5.2 in \cite{Neri:2025fsh} for a summary). In this text, we will set aside the challenges stemming from the lack of a comprehensive general theory and assume we have knowledge of the specific orbits and quantization methods that yield the unitary representations of interest.

From the perspective of the orbit method, the classical counterpart of a spin chain may be interpreted as a mechanical system whose phase space is a product of specific coadjoint orbits. Conversely, the phase space of a particle on $\mathcal{M}$ is the cotangent bundle $\mathrm{T}^\ast\mathcal{M}$. Assuming a correspondence between such phase spaces and their associated Hamiltonians,
\begin{equation}\label{ClassicalRelation}
\Big(\mathrm{T}^\ast\mathcal{M}\,, H_{\text{free particle}}\Big) \qquad \leftrightsquigarrow \qquad \Big(\mathcal{O}_1 \times \mathcal{O}_2 \times \ldots \times \mathcal{O}_n\,, H_{\text{spin}}\Big),
\end{equation}
for certain coadjoint orbits $\mathcal{O}_i$ and a spin-chain Hamiltonian $H_{\text{spin}}$ on their product, we aim to replace the complex object of quantization -- the cotangent bundle with the geodesic Hamiltonian -- with a product of simpler symplectic manifolds, namely coadjoint orbits, equipped with a ``spin'' Hamiltonian. Quantizing this correspondence then leads to a relation of the form
\begin{equation}
\Big(L^2(\mathcal{M})\,, -\triangle\Big) \qquad \leftrightsquigarrow \qquad \Big(\mathsf{V}_1 \otimes \mathsf{V}_2 \otimes \ldots \otimes \mathsf{V}_n\,, H_{\text{spin}}\Big)\,,
\end{equation}
where each Hilbert space $\mathsf{V}_i$ is obtained via geometric quantization of the orbit $\mathcal{O}_i$. In an ideal scenario, this correspondence would imply the equivalence of the respective spectral problems.
More specifically, we shall discuss the following principle\footnote{This work stands as a further testament to the ``Symplectic Creed'' \cite{Weinstein1981}: \textit{Everything is a Lagrangian Submanifold}.} \cite{Bykov:2012am}: 

\begin{quote}
\textit{If there exists a Lagrangian embedding of the particle's configuration space into a product of coadjoint orbits, which in a suitable sense extends to a symplectomorphism of the particle and spin chain phase spaces, then a spin Hamiltonian having a minimum on this Lagrangian submanifold induces the dynamics of a free particle on that Lagrangian submanifold, endowed with a metric given by the quadratic momentum part of the spin Hamiltonian.}
\end{quote}

Providing full mathematical rigor for these ideas is challenging; in the main body of the text, we attempt to formalize them, at least in part. Although a completely general theory for this type of reasoning is currently lacking, there exist several concrete examples where such a correspondence can be rigorously established.

We now present a brief history of ideas relevant to our survey. It begins with the work of D. Haldane\footnote{This insight formed a significant part of Haldane's Nobel Prize contribution; see Section 5 of the \href{https://www.nobelprize.org/uploads/2018/06/advanced-physicsprize2016.pdf}{Scientific Background on the Nobel Prize in Physics 2016}. For a modern exposition, see the review \cite{Affleck:2021vzo} and references therein.}, who showed that the low-energy effective theory for antiferromagnetic $\mathbf{SU}(2)$ infinite spin chain in the large-spin limit is the seminal $(1+1)$-dimensional $\mathbf{O}(3)$ nonlinear sigma model, modified by a topological $\theta$-term \cite{Haldane:1983ru,HALDANE1983464}. A mathematical interpretation of this construction through symplectic and Lagrangian geometry was later proposed by D. Bykov \cite{Bykov:2011ai,Bykov:2012am}; in Section \ref{SectionGeneral}, we follow the ideas from the second of these papers.

Subsequently, attention shifted to the more tractable and mathematically rigorous one-dimensional case. In\footnote{For a short summary see \cite{Kuzovchikov:2025mwc}.} \cite{Bykov:2024tvb}, D. Bykov and A. Kuzovchikov constructed $\mathbf{SU}(n)$ ``all-to-all'' $n$-site spin chains, whose large-spin limit yields one-dimensional sigma models on flag manifolds; we review this construction in Sections \ref{SectionSphere} and \ref{FlagSection}. In the same work and its sequel \cite{BykovGaudin}, they further demonstrated how algebraic methods can be used to obtain the spectrum of the Laplace–Beltrami operator on flag manifolds endowed with various invariant metrics, solving, for example, the general case of the complete flag $\mathcal{F}_3$. The paper \cite{Bykov:2025wvn} takes initial steps for isotropic flag manifolds by constructing several relevant Lagrangian embeddings into $\mathbf{SO}$ and $\mathbf{Sp}$ coadjoint orbits; in collaboration with the author, the work \cite{Bykov:2024wft} also introduces $\mathcal{N} = 2$ and $\mathcal{N} = 4$ supersymmetric $\mathbf{SU}(n)$ spin chains, whose continuum limits correspond to supersymmetric one-dimensional flag sigma models.

Sections \ref{FlatSection} and \ref{HyperbolicSection} constitute new examples to the subject. In these, we quantize a particle on $\mathbb{C}$ as a ``Heisenberg chain'' built from two ladder operators algebras, and provide a physical interpretation of the mathematical fact that the tensor product of positive and negative discrete series representations of $\mathbf{SL}(2,\mathbb{R})$ is isomorphic to the space of square-integrable functions on the hyperbolic plane \cite{Repka}. In a subsequent joint paper~\cite{JointPaper} we aim to provide a more in-depth analysis of these examples.

It would also be interesting to develop the theory discussed here in broader contexts, for instance for infinite groups -- such as loop groups, the BMS group, or the Virasoro–Bott group \cite{Khesin2009TheGO} -- as well as for field theories and infinite spin chains, particles on (anti-)de Sitter-type spaces, among others. It would also be instructive to establish connections with other geometrical approaches to quantization, such as the WKB quantization of Lagrangian manifolds \cite{BatesWeinstein1997} and the mixed classical-quantum formalism \cite{Liashyk:2024ekz}. Another promising direction is to investigate potential applications of further Lagrangian embeddings, such as those discussed in \cite{tyurin2025ample,Tyurin2021,tyurin2025examples,mironov2004new}.

\section{From spin chains to sigma models: general idea}\label{SectionGeneral}
\subsection{Classical part}\label{SectionGeneralClassical}
Suppose we wish to establish a relation of the form \eqref{ClassicalRelation} between a free particle on a Riemannian manifold\footnote{For simplicity, we assume in this Section that the particle is free from interactions with additional structures such as magnetic fields or fermions.} $\mathcal{M}$ with metric $g$ and a classical spin chain whose phase space is a product $\mathcal{O}_1 \times \mathcal{O}_2 \times \ldots \times \mathcal{O}_n$ of certain coadjoint orbits, governed by a suitable Hamiltonian. In this Section, we present a method that applies under specific assumptions.

The first assumption is the existence of a Lagrangian embedding
\begin{equation}\label{GeneralLagrEmb}
\mathcal{M}\hookrightarrow\mathcal{O}_1\times\mathcal{O}_2\times\ldots\times \mathcal{O}_n
\end{equation}
with respect to the sum of the Kirillov-Kostant-Souriau forms.
If such an embedding exists, it relates the kinematics of a point particle to that of a classical spin chain in the following way. By the seminal \textit{Darboux--Weinstein theorem} \cite{WEINSTEIN1971329}, there exist an open neighborhood $U(\mathrm{T}^\ast\mathcal{M})$ of the zero section in the cotangent bundle of $\mathcal{M}$ and an open neighborhood $U(\mathcal{O}_1\times\mathcal{O}_2\times\ldots\times \mathcal{O}_n)$ of the Lagrangian image of $\mathcal{M}$ such that the two neighborhoods are symplectomorphic:
\begin{equation}\label{ZeroSectionNeighborhoods}
U(\mathrm{T}^\ast\mathcal{M})\simeq U(\mathcal{O}_1\times\mathcal{O}_2\times\ldots\times \mathcal{O}_n)\,.
\end{equation}
The second assumption is that $U(\mathcal{O}_1\times\mathcal{O}_2\times\ldots\times \mathcal{O}_n)$ is ``suitable dense'' in the product of coadjoint orbits. Morally, this means that
\begin{equation}\label{LargeU}
U(\mathcal{O}_1\times\mathcal{O}_2\times\ldots\times \mathcal{O}_n)\simeq\frac{\mathcal{O}_1\times\mathcal{O}_2\times\ldots\times \mathcal{O}_n}{\text{some ``small'' closed submanifolds}}\,. \end{equation}
As we will see below, $U(\mathrm{T}^\ast\mathcal{M})$ can likewise be made to approximate the entire cotangent bundle $\mathrm{T}^\ast\mathcal{M}$ simply by rescaling the momenta, which live in the linear fibers. Thus, the core idea is that the existence of the Lagrangian embedding \eqref{GeneralLagrEmb} can lead to an ``almost complete'' matching between the kinematics of a particle on the manifold and that of the spin chain.

Let us now turn to the dynamical aspect of the correspondence. Specifically, how can the dynamics on the product of orbits induce dynamics on the Lagrangian submanifold? Denote the Liouville 1-form on the product of orbits by $\theta_\lambda = \lambda\theta$, where $\lambda\in\mathbb{R}$ is a formal scale parameter with respect to a reference 1-form $\theta$.
A further ingredient in our construction is a Hamiltonian $\lambda^2 H_{\text{spin}}$ that possesses a global minimum exactly on the Lagrangian submanifold $\mathcal{M}$. Without loss of generality, we may set the value of $H_{\text{spin}}$ on $\mathcal{M}$ to zero. The classical action for the spin chain is then
\begin{equation}\label{GeneralClassicalSpinChain}
\mathcal{S} = \int\Big(\theta_\lambda - \lambda^2H_{\text{spin}}\dif t\Big)\,.
\end{equation}
Due to the ``almost symplectomorphism'' \eqref{ZeroSectionNeighborhoods}, we can introduce canonical coordinates $q^i$ and momenta $p_i$ on the product of orbits, induced from the neighborhood $U(\mathrm{T}^\ast\mathcal{M})$. Moreover, since $H_{\text{spin}}(q,p)$ attains its minimum on $\mathcal{M}$, defined by the zero section $\{p_i = 0\}$, we can expand it in a Taylor series in the momenta, beginning with a quadratic term. Consequently, the action \eqref{GeneralClassicalSpinChain} takes the form
\begin{equation}
\mathcal{S}[p,q] = \int\dif t\bigg(\lambda p_i\dot{q}^i + \frac{\lambda^2 g^{ij}(q) p_ip_j}{2}+O\big(\lambda^2p^3\big)\bigg)\,,
\end{equation}
where $g^{ij}(q)$ denotes the coefficients of the quadratic part in the expansion of the spin Hamiltonian. The second term in the action then resembles the geodesic Hamiltonian of a free point particle with the configuration space $\mathcal{M}$,
\begin{equation}\label{GeneralFreeHamiltonian}
H_{\text{free particle}} = \frac{g^{ij}(q) p_ip_j}{2}\,.
\end{equation}

However, two obstacles prevent us from directly identifying this system with a point particle. First, we have correction terms of order $O\big(\lambda^2p^3\big)$. Second, the momenta $p$ belong only to the neighborhood $U(\mathrm{T}^\ast\mathcal{M})$ rather than the full cotangent bundle $\mathrm{T}^\ast\mathcal{M}$. Both issues can be resolved simultaneously by the following rescaling trick: let us rescale the momenta as $p\rightarrow\lambda^{-1}p$ and then take the limit $\lambda\rightarrow\infty$. In this limit, $U(\mathrm{T}^\ast\mathcal{M})$ expands to the entire $\mathrm{T}^\ast\mathcal{M}$, and the higher-order corrections $O\big(\lambda^{-1}p^3\big)$ are suppressed. Consequently, we recover the standard first-order action for a point particle\footnote{Note that the geometric parameter $\lambda$ plays distinct roles for the free particle and for the spin chain. In the classical spin chain, $\lambda$ scales the symplectic form, whereas for the free particle it determines the size of the Darboux–Weinstein neighbourhood.} on $\mathcal{M}$:
\begin{equation}
\mathcal{S} = \int\dif t\bigg( p_i\dot{q}^i+\frac{g^{ij}(q) p_ip_j}{2}\bigg)\simeq\frac{1}{2}\int\dif t\big(g_{ij}\dot{q}^{\,i}\dot{q}^{\,j}\big)\,.
\end{equation}
This also implies that the kinematical relation, in the so-called ``large spin'' limit\footnote{If $U(\mathrm{T}^\ast\mathcal{M})$ is already symplectomorphic to $\mathrm{T}^\ast\mathcal{M}$ and the Hamiltonian is purely quadratic, then taking the limit $\lambda\rightarrow\infty$ is unnecessary. For uniformity we retain the notation, bearing in mind that it may sometimes be omitted.} $\lambda\rightarrow\infty$, becomes \begin{equation}\label{GeneralAlmostSymplectomorphism2} \mathrm{T}^\ast\mathcal{M}\,\,\text{``$\simeq$''}\,\lim_{\lambda\rightarrow\infty}\mathcal{O}_1\times\mathcal{O}_2\times\ldots\times \mathcal{O}_n\,.
\end{equation}

Thus, the recipe for rewriting a classical point particle as a classical spin chain can be summarized as follows:
\begin{enumerate}
\item Find a Lagrangian embedding \eqref{GeneralLagrEmb} and use it to identify the corresponding neighborhoods of the zero sections, establishing the symplectomorphism \eqref{ZeroSectionNeighborhoods}. Ensure the neighborhood in the product of orbits is sufficiently large, see \eqref{LargeU}.
\item Choose a Hamiltonian on the product of orbits that attains a global minimum on the Lagrangian submanifold and whose quadratic part \eqref{GeneralFreeHamiltonian} in the Taylor expansion defines the desired metric.
\end{enumerate}

\subsection{Quantization}\label{GeneralQuantSection}
In the previous Section, we established a connection between point particles on Riemannian manifolds and spin chains in the classical setting. We now turn to its quantum counterpart.

Quantizing the relation \eqref{GeneralAlmostSymplectomorphism2} under the assumptions outlined above\footnote{Geometric quantization typically imposes integrality conditions on $\lambda$. For simplicity, we assume normalizations such that $\lambda\in\mathbb{Z}_{\geq 0}$.} yields the quantum kinematical relation
\begin{equation}\label{L2AsSpinChain}
L^2(\mathcal{M}) \simeq \lim_{\lambda\rightarrow\infty}\big(\mathsf{V}^\lambda_1\otimes\mathsf{V}^\lambda_2\otimes\ldots\otimes \mathsf{V}^\lambda_n\big)\,,
\end{equation}
where each representation $\mathsf{V}^\lambda_i$ is obtained by the geometric quantization of the coadjoint orbit $\mathcal{O}_i$. Here, $\lambda$ plays the role of a representation parameter, such as spin. We assume that the ``small submanifolds'' excluded in \eqref{LargeU} do not significantly affect the geometric quantization process.

Note that establishing this isomorphism explicitly can be subtle. Typically, each $\mathsf{V}^\lambda_i$ is realized as the space of polarized sections with respect a distribution $\mathscr{P}$ of a certain line bundle $\mathscr{E}_i$ over $\mathcal{O}_i$, see \cite{Wernli:2023pib,Kirillov2001}. The desired isomorphism then amounts to constructing a map\footnote{The bundle $\mathscr{E}_1\boxtimes\mathscr{E}_2\boxtimes\ldots\boxtimes\mathscr{E}_n$ is typically non-trivial over the product of orbits, but it typically admits a trivialization over the dense open subset $\mathcal{O}_1\times\mathcal{O}_2\times\ldots\times \mathcal{O}_n/\mathscr{D}$. This trivialization is essential for identifying sections with functions, as illustrated in Section \ref{SectionSphere}.}
\begin{equation} \Gamma(\mathscr{E}_1\boxtimes\mathscr{E}_2\boxtimes\ldots\boxtimes\mathscr{E}_n,\mathscr{P})\longrightarrow C^\infty(\mathcal{M}) 
\end{equation}
from the space of polarized sections of the external tensor product bundle to smooth functions on $\mathcal{M}$. Constructing such a map generally involves several steps: a suitable trivialization of the bundle over the dense open set $\mathcal{O}_1\times\mathcal{O}_2\times\ldots\times \mathcal{O}_n/\mathscr{D}$, where $\mathscr{D}$ is the excluded ``small'' submanifold in \eqref{LargeU}, a change of polarization, and restriction (pullback) of the bundle to the Lagrangian submanifold $\mathcal{M}$.

The dynamical relation expected after quantization takes the form
\begin{equation}
\mathsf{Spec}(-\triangle)\simeq\lim_{\lambda\rightarrow\infty}\mathsf{Spec}_\lambda(H_{\text{spin}})\,,
\end{equation}
where $\mathsf{Spec}$ denotes the spectrum (i.e., the set of eigenvalues) of the operator acting on $L^2(\mathcal{M})$, and $\mathsf{Spec}_\lambda$ denotes the spectrum\footnote{Strictly speaking, the large $\lambda$ limit requires careful functional-analytic treatment, for instance in terms of spectral measures. However, in all known examples (see the Section \ref{SectionSphere}) the inclusion $\mathsf{Spec}_\lambda(H_{\text{spin}})\subset\mathsf{Spec}_{\lambda+1}(H_{\text{spin}})$ holds, so the limit can be understood as a direct limit in the set-theoretic sense.} of an operator acting on $\mathsf{V}^\lambda_1\otimes\ldots\otimes \mathsf{V}^\lambda_n$. This relation provides a powerful new tool for computing the spectra of $-\triangle$ on complicated Riemannian manifolds.

\section{Examples}
While the general procedure of mapping a spin chain to a particle on a manifold may appear rather involved and suffers on numerous assumptions, this Section aims to illustrate the construction through a series of concrete examples.
\subsection{Particle on $\mathbb{C}$ as two oscillators}\label{FlatSection}
As a warm-up example, let us consider a free particle on the complex plane. Its action is simply
\begin{equation}
\mathcal{S}[q] = \frac{1}{2}\int\dif t\,|\dot{q}|^2\,,
\end{equation}
where $q$ is a complex coordinate on $\mathbb{C}$. We can rewrite this action in first-order form by introducing a complex momentum $p$:
\begin{equation}\label{1stOrderPlane}
\mathcal{S}[p,q] = \frac{1}{2}\int\dif t\Big(i\smallthickbar{p}\dot{q}-ip\dot{\smallthickbar{q}} - |p|^2\Big)\,.
\end{equation}
Now perform a linear change of variables
\begin{equation}
p = z-\smallthickbar{w}\,,\qquad q = z+\smallthickbar{w}\,,
\end{equation}
where $z$ and $w$ are new complex coordinates. This transformation diagonalizes the kinetic term in \eqref{1stOrderPlane}. The action then becomes equivalent to
\begin{equation}\label{ActionZWplane}
\mathcal{S}[z,w] = \int\dif t\Big(i\smallthickbar{z}\dot{z}+i\smallthickbar{w}\dot{w} - |z-\smallthickbar{w}|^2\Big)\,.
\end{equation}

This action admits a natural interpretation within our framework. Note that the particle's configuration space $\mathbb{C}_{\text{conf}} \simeq \{p = 0\} = \{z = \smallthickbar{w}\}$ is embedded into $\mathbb{C} \times \mathbb{C}$ as a Lagrangian submanifold with respect to the symplectic form $\omega = i\dif\smallthickbar{z}\wedge\dif z + i\dif\smallthickbar{w}\wedge\dif w$. Moreover, the Hamiltonian $H = |z - \smallthickbar{w}|^2$ on $\mathbb{C} \times \mathbb{C}$ attains a global minimum precisely on $\mathbb{C}_{\text{conf}}$. Finally, $\mathbb{C}$ viewed as a phase space can be identified with a coadjoint orbit inside the Heisenberg algebra\footnote{The complex planes $\mathbb{C}$ can even be viewed as centrally extended $\mathbf{ISO}(2)$ coadjoint orbits.}, see the Section 4.3 in \cite{Lahlali:2024qnk}. Consequently, following the general recipe\footnote{In this particular example, the formal ``large‑scale limit'' $\lambda \to \infty$ and exclusion of some ``small'' submanifolds are not required.} of Section \ref{SectionGeneralClassical}, we have rewritten a particle on the plane as a two‑site, non‑compact classical ``spin chain'' associated with the Heisenberg group.

The quantization of the model \eqref{ActionZWplane} is straightforward. Owing to the symplectic structure, the coordinates $z$ and $w$ are quantized as two independent oscillators $a$ and $b$, satisfying the standard commutation relations $[a,a^\dagger]=1$ and $[b,b^\dagger]=1$. It is instructive to represent them in the Bargmann–Fock space $\mathscr{B}\big(\mathbb{C}^2\big)$ by
\begin{equation}
a^\dagger = z\,,\qquad a=\frac{\partial}{\partial z}\,,\qquad b^\dagger = w\,,\qquad b = \frac{\partial}{\partial w}\,,\qquad z,w\in\mathbb{C}\,.
\end{equation}
Wave functions in this representation are entire functions in $z$ and $w$, and the Hilbert‑space measure is given by
\begin{equation}\label{BFflatWaves}
\dif\mu = e^{-|z|^2-|w|^2}\dif^2 z\,\dif^2 w\,.
\end{equation}
The eigenvalues and eigenfunctions of the Hamiltonian $H = \big(a - b^\dagger\big)\big(a^\dagger - b\big)$ are
\begin{equation}
\Psi_{\alpha}(z,w) = \exp\big({zw + \alpha z - \smallthickbar{\alpha} w}\big)\,,\qquad H\Psi_\alpha = |\alpha|^2\Psi_\alpha\,,
\end{equation}
which reproduces the continuous spectrum of a free particle on the plane. To recover the standard plane waves, it is convenient to embed $\mathscr{B}\big(\mathbb{C}^2\big)$ isometrically into $L^2\big(\mathbb{C}^2\big)$ via the map
\begin{equation}
\Psi(z,w)\longmapsto \widetilde{\Psi}(z,\smallthickbar{z},w,\smallthickbar{w}) = e^{-\frac{1}{2}|z|^2-\frac{1}{2}|w|^2}\Psi(z,w)\,.
\end{equation}
Now, on $L^2\big(\mathbb{C}^2\big)$ the restriction to the Lagrangian submanifold $\{\smallthickbar{z}=w\}$ is well defined, yielding
\begin{equation}
\widetilde{\Psi}_\alpha(z,\smallthickbar{z},w,\smallthickbar{w})\Big|_{w=\smallthickbar{z}} = e^{i\,\mathrm{Im}(\alpha z)}\qquad-\qquad\text{plane waves on }\mathbb{C}\,.
\end{equation}

\subsection{Particle on $\mathbb{S}^2$ and $\mathbf{SU}(2)$ spin chain}\label{SectionSphere}
A more non-trivial example is the representation of a free particle on the two-dimensional sphere $\mathbb{S}^2\simeq\mathbb{CP}^1$ as a two-site $\mathbf{SU}(2)$ XXX spin chain in the large-spin limit. Following the general approach outlined in the previous Section, the starting point is a suitable Lagrangian embedding of the sphere into a product of coadjoint orbits. Since $\mathbf{SU}(2)$ is the isometry group of the sphere, it is natural to consider its coadjoint orbits. All coadjoint orbits of $\mathbf{SU}(2)$ are themselves spheres, equipped with a symplectic form proportional to the volume form. These spheres are conveniently parametrized by complex vectors in $\mathbb{C}^2$ of unit length, identified up to a phase (``Hopf fibration parameterization'').

We therefore take as the classical spin-chain phase space the product $\mathbb{CP}^1\times\mathbb{CP}^1$, endowed with the symplectic form
\begin{equation}\label{TwoSpheresSymplecticForm}
\omega_\lambda = i\lambda\big(\dif\smallthickbar{z}\wedge\dif z+\dif\smallthickbar{w}\wedge\dif w\big)\,,\qquad |z|^2=|w|^2 = 1\,,
\end{equation}
where the two-component complex vectors $z$ and $w$ parametrize the first and second sphere, respectively, and summation over $\mathbb{C}^2$ indices is implied. The following Lagrangian embedding is central to the construction.

\begin{quote}
\underline{\textbf{Statement.}} The submanifold of $\big(\mathbb{CP}^1\times\mathbb{CP}^1,\omega_\lambda\big)$ defined by the equation\footnote{Here and hereafter, for vectors $u,v\in\mathbb{C}^n$, we denote by $\smallthickbar{v}\cdot u$ the standard Hermitian inner product.} $\smallthickbar{z}\cdot w = 0$ is Lagrangian and diffeomorphic to a sphere.
\end{quote}

This claim can be readily verified in the particular parameterization of the equation $\smallthickbar{z}\cdot w = 0$ as $w_i = \varepsilon_{ij}\smallthickbar{z}^{\,j}$, where $\varepsilon_{ij}$ is the two-dimensional Levi-Civita symbol. It is also useful to understand this embedding in the inhomogeneous coordinates. Parametrizing $z = [1,z_{\text{inh}}]$ and $w=[w_{\text{inh}},-1]$, the condition $\smallthickbar{z}\cdot w = 0$ becomes $z_{\text{inh}} = \smallthickbar{w}_{\text{inh}}$, which embeds the sphere into the product as $z_{\text{inh}}\mapsto(z_{\text{inh}},\smallthickbar{z}_{\text{inh}})$.

As discussed, the Darboux–Weinstein theorem suggests that such a Lagrangian embedding can be extended to a symplectomorphism of the type \eqref{GeneralAlmostSymplectomorphism2}. The precise statement is as follows.

\begin{quote}
\underline{\textbf{Theorem.}} (\cite{Bykov:2024tvb}) There exists a symplectomorphism\footnote{The statement concerning the large $\lambda$ limit should be understood as follows: at finite $\lambda$, we obtain a symplectomorphism onto a Darboux-Weinstein tubular neighborhood $U_\lambda\big(\mathrm{T}^\ast\mathbb{CP}^1\big)$ whose linear size scales with $\lambda$. Taking $\lambda\rightarrow\infty$ enlarges this neighborhood to the entire $\mathrm{T}^\ast\mathbb{CP}^1$.}
\begin{equation}\label{TheorCotB}
\mathrm{T}^\ast\mathbb{CP}^1\simeq\lim_{\lambda\rightarrow\infty}\frac{\big(\mathbb{CP}^1\times\mathbb{CP}^1,\omega_\lambda\big)}{\{\det\mathcal{Z}=0\}}\,,
\end{equation}
where $\mathcal{Z}$ is the $2\times 2$ matrix whose columns are the vectors $z$ and $w$.
\end{quote}

We now sketch the proof and explain why the large $\lambda$ limit is required and why the determinantal variety $\mathscr{D}\simeq\{\det\mathcal{Z}=0\}$ must be removed. Informally, the coordinates on $\mathrm{T}^\ast\mathbb{CP}^1$ are obtained from $\big(\mathbb{CP}^1\times\mathbb{CP}^1,\omega_\lambda\big)$ via a polar decomposition of $\mathcal{Z}$ in the following way. First, rescale the coordinates as $z\rightarrow\lambda^{-\frac{1}{2}}z$ and $w\rightarrow\lambda^{-\frac{1}{2}}w$. The Gram matrix $G:=\mathcal{Z}^\dagger\mathcal{Z}$ of inner products then takes the form
\begin{equation}
G = \begin{pmatrix}
\lambda & \smallthickbar{z}\cdot w\\
\smallthickbar{w}\cdot z & \lambda
\end{pmatrix}\,,
\end{equation}
where we have used the rescaled normalization $|z|^2 = |w|^2 = \lambda$. The idea is to introduce a new ``momentum'' variable $K:=\smallthickbar{z}\cdot w$. However, the Cauchy–Schwarz inequality imposes the constraint $\lambda^2\geq |K|^2$, which is satisfied for all finite $K$ only in the limit $\lambda\rightarrow\infty$. Moreover, provided $\det\mathcal{Z}\neq0$, there exists a unique polar decomposition 
\begin{equation}\label{PolarDecomposition}
\mathcal{Z} = U\sqrt{G}\,,
\end{equation}
where $U$ is an $\mathbf{SU}(2)$ matrix that, up to an irrelevant phase $e^{i\varphi J_3}$, can be parametrized as (see formula (6.51) in \cite{BengtssonZyczkowski2017})
\begin{equation}
U(q) = e^{q J_-}e^{-\ln(1+|q|^2)J_3}e^{-\smallthickbar{q}J_+} = \frac{1}{\sqrt{1+|q|^2}}\begin{pmatrix}
1 & -\smallthickbar{q}\,\\
q & 1
\end{pmatrix}\,,\qquad q\in\mathbb{C}\,.
\end{equation}
Here $J_\pm$ and $J_3$ denote the Cartan basis of the fundamental representation of $\mathfrak{su}(2)$.

The theorem essentially asserts that the map $(z,w)\mapsto(q,K)$ gives a coordinate transformation to $\mathrm{T}^\ast\mathbb{CP}^1$, with $K$ playing the role of a (complex) momentum and $q$ being the (complex) stereographic coordinate on the sphere. The remaining part of the proof consists in verifying directly that this map is indeed a symplectomorphism. Rewriting the symplectic form \eqref{TwoSpheresSymplecticForm} in terms of $\mathcal{Z}$ as $\omega_\lambda = i\,\mathsf{Tr}\big(\dif\mathcal{Z}^\dagger\wedge\dif\mathcal{Z}\big)$ and substituting the decomposition \eqref{PolarDecomposition}, a straightforward computation yields
\begin{equation}
\omega_\lambda = i\,\dif\mathsf{Tr}\big(G\,U^\dagger\dif U\big) = i\,\dif\bigg(\frac{K\dif q}{1+|q|^2}\bigg)\,,
\end{equation}
where we have used the fact that the sphere is embedded in its two copies as a Lagrangian submanifold. Introducing the canonical momentum $p = \big(1+|q|^2\big)^{-1}K$ finally produces the Darboux coordinates induced from $\mathrm{T}^\ast\mathbb{CP}^1$, which completes the proof. $\hfill\ensuremath{\Box}$ 

We have covered the kinematics; it is now time to introduce a Hamiltonian into the game. Recall that the general prescription requires a real function on $\mathbb{CP}^1\times\mathbb{CP}^1$ that possesses a minimum on the Lagrangian sphere $\{\smallthickbar{z}\cdot w = 0\}$. The simplest candidate for this role is $H = |\smallthickbar{z}\cdot w|^2$. This is a natural generalization of the Hamiltonian from Section \ref{FlatSection}, as it can be expressed in inhomogeneous coordinates in the form
\begin{equation}
H = \frac{|\smallthickbar{z}_{\text{inh}} - w_{\text{inh}}|^2}{\big(1+|z_{\text{inh}}|^2\big)\big(1+|w_{\text{inh}}|^2\big)}\,.
\end{equation}
Thus, we arrive at the classical spin-chain action in the rescaled variables:
\begin{equation}\label{SphereSpinChainClassical}
\mathcal{S}[z,w] = \int\Big(i\smallthickbar{z}\cdot\dot{z}+i\smallthickbar{w}\cdot\dot{w}-|\smallthickbar{z}\cdot w|^2\Big)\dif t\,,\qquad |z|^2=|w|^2 = \lambda\,.
\end{equation}
It turns out that this simplest choice is in fact the best one. One can demonstrate, through a direct change of variables $(z,w)\mapsto(q,p)$ and elimination of $p$ via the equations of motion in the limit $\lambda\rightarrow\infty$, that the action \eqref{SphereSpinChainClassical} is equivalent to the sphere sigma model action
\begin{equation}
\mathcal{S}[q] = \int\dif t\,\Bigg(\frac{\dot{q}\dot{\smallthickbar{q}}}{\big(1+|q|^2\big)^2}\Bigg)\,.
\end{equation}
The calculations needed to verify this equivalence follow almost identically the reasoning provided in the proof of the theorem above.

We now proceed to canonical quantization of the theory \eqref{SphereSpinChainClassical}. The symplectic structure of the kinetic term leads to the quantization of $z$ and $w$ into two two-component ladder operators $a$ and $b$ satisfying the standard commutation relations
\begin{equation}
[a_i,a^\dagger_j] = \delta_{ij}\,,\qquad [b_i,b^\dagger_j] = \delta_{ij}\,.
\end{equation}
Consequently, the Hilbert space is the Fock space for four oscillators. The constraints on the vectors $|z|^2=|w|^2 = \lambda$ are quantized as constraints on the occupation numbers in the Fock space:
\begin{equation}\label{QuantumSphereConstraint}
a^\dagger\cdot a:=a^\dagger_1 a_1+a^\dagger_2 a_2 = \lambda\mathds{1}\,,\qquad b^\dagger\cdot b:=b^\dagger_1 b_1+b^\dagger_2 b_2 = \lambda\mathds{1}\,,
\end{equation}
where $\mathds{1}$ is the identity operator. This leads to the quantization condition $\lambda\in\mathbb{Z}_{\geq 0}$. The same condition can also be derived in the framework of geometric quantization by requiring that $\exp \big(i\int\theta_\lambda\big)$ is globally well defined, see \cite{Alvarez:1984es,Affleck:2021vzo}. In fact, it is easy to see that the Fock space constrained by \eqref{QuantumSphereConstraint} is isomorphic to $\mathsf{V}_{\frac{\lambda}{2}}\otimes\mathsf{V}_{\frac{\lambda}{2}}$, where $\mathsf{V}_{\frac{\lambda}{2}}$ is the spin-$\frac{\lambda}{2}$ representation of $\mathfrak{su}(2)$. This agrees with the general fact that the geometric quantization of a sphere of radius $\lambda$ yields precisely the representation $\mathsf{V}_{\frac{\lambda}{2}}$. This representation is realized as the space of holomorphic global sections of the holomorphic line bundle $\mathscr{O}(\lambda)$ over the sphere. Specifically, these sections are homogeneous polynomials of degree $\lambda$ in the homogeneous coordinates of $\mathbb{CP}^1$, on which $\mathbf{SU}(2)$ acts irreducibly via its natural matrix multiplication on the vector of homogeneous coordinates.

Finally, the Hamiltonian is quantized as the operator $H = \big(a^\dagger \cdot b\big)\big(b^\dagger \cdot a\big)$, which can be identified (up to an additive constant) with the XXX Hamiltonian
\begin{equation}
H = \frac{1}{2}\sum\limits_{m = 1}^3 S^m\otimes S^m
\end{equation}
via the so-called \textit{Schwinger-Wigner map} \cite{Affleck:2021vzo}
\begin{equation}
S^m\otimes\mathds{1} = a^\dagger\cdot\sigma^m\cdot a := a^\dagger_i\sigma^m_{ij} a_j\,,\qquad \mathds{1}\otimes S^m = b^\dagger\cdot\sigma^m\cdot b:=b^\dagger_i\sigma^m_{ij} b_j\,.
\end{equation}
Here $S^m$ denotes the $m$'th generator of $\mathfrak{su}(2)$ in the $\mathsf{V}_{\frac{\lambda}{2}}$ representation, and $\sigma^m$ are the Pauli matrices. This identification can be verified using the formula
\begin{equation}
\sum\limits_{m = 1}^3\sigma^m\otimes\sigma^m = 2\mathbb{P}_{12} - \mathds{1}\,,\qquad \mathbb{P}_{12}(v\otimes u):= u\otimes v\,,
\end{equation}
since
\begin{equation}
\frac{1}{2}\sum\limits_{m = 1}^3 S^m\otimes S^m = \big(a^\dagger\otimes b^\dagger\big)\Bigg(\sum\limits_{m = 1}^3\sigma^m\otimes\sigma^m\Bigg)\big(a\otimes b\big) = \big(a^\dagger \cdot b\big)\big(b^\dagger \cdot a\big) - \frac{\lambda^2}{2}\,,
\end{equation}
where we used the constraints \eqref{QuantumSphereConstraint}.

Why is all this stuff useful? Observe that the Hamiltonian $H = \big(a^\dagger \cdot b\big)\big(b^\dagger \cdot a\big)$ acts on the Hilbert space
\begin{equation}
\mathscr{H}\simeq \mathsf{V}_{\frac{\lambda}{2}}\otimes\mathsf{V}_{\frac{\lambda}{2}}\simeq\bigoplus\limits_{k = 0}^\lambda\mathsf{V}_k\,.
\end{equation}
It can be diagonalized explicitly by eigenfunctions of the form
\begin{equation}\label{EigenfunctionsSphere}
\Psi_k = T_{i_1 i_2\ldots i_k j_1j_2\ldots j_k}a^\dagger_{i_1}a^\dagger_{i_2}\ldots a^\dagger_{i_k}b^\dagger_{j_1}b^\dagger_{j_2}\ldots b^\dagger_{j_k}\big(\varepsilon_{ij}a^\dagger_i b^\dagger_j\big)^{\lambda - k}|0\rangle\,,
\end{equation}
where $T$ is a symmetric rank $2k$ tensor realizing the $\mathsf{V}_k$ representation, $\varepsilon_{ij}$ is the fully antisymmetric tensor, and summation over repeated indices is implied. The corresponding eigenvalues are $H\Psi_{k} = k(k+1)\Psi_k$ with $k=0,\ldots,\lambda$. This coincides precisely with the spectrum of the Laplace–Beltrami operator on the sphere, truncated at the level of the first $\lambda$ harmonics. Therefore, in full accordance with the general idea, the limit $\lambda\rightarrow\infty$ recovers the complete spectrum of $-\triangle$ on the sphere, obtained through purely algebraic means rather than by solving a second‑order elliptic partial differential equation.

The method also reconstructs the spherical harmonics themselves. The eigenfunctions \eqref{EigenfunctionsSphere} can be realized as sections of the holomorphic line bundle $\mathscr{O}(\lambda)\boxtimes\mathscr{O}(\lambda)$ over $\mathbb{CP}^1\times\mathbb{CP}^1$:
\begin{equation}
\Psi_k(z,w) = T_{i_1\ldots i_k j_1\ldots j_k} z^{i_1}\ldots z^{i_k} w^{j_1}\ldots w^{j_k}\big(\det\mathcal{Z}\big)^{\lambda - k}\,,\qquad \det\mathcal{Z}:=\varepsilon_{ij}z^i w^j\,.
\end{equation}
A potential difficulty is that $\Psi_k$ is a section of a bundle over the product of orbits, while spherical harmonics are functions on the Lagrangian sphere. The resolution lies in the fact that, according to \eqref{TheorCotB}, we are actually interested in the restriction of the bundle $\mathscr{O}(\lambda)\boxtimes\mathscr{O}(\lambda)$ to the open dense subset $\mathscr{X}:=\big(\mathbb{CP}^1\times\mathbb{CP}^1\big)/\mathscr{D}$, where the divisor $\mathscr{D}$ is the zero locus of $\det\mathcal{Z}$. By construction, $\big(\det\mathcal{Z}\big)^\lambda$ is a nowhere‑vanishing section of $\mathscr{O}(\lambda)\boxtimes\mathscr{O}(\lambda)$ over $\mathscr{X}$. Consequently, this section trivializes the bundle over $\mathscr{X}$; every section $s\in\Gamma\big(\mathscr{O}(\lambda)\boxtimes\mathscr{O}(\lambda),\mathscr{X}\big)$ can be written as $s(z,w) = f(z,w)\big(\det\mathcal{Z}\big)^\lambda$, where $f(z,w)$ is an ordinary function on $\mathscr{X}$. The final step is to restrict $f(z,w)$ to the Lagrangian submanifold defined by $\smallthickbar{z}\cdot w = 0$.

\begin{quote}
\underline{\textbf{Statement.}} The functions
\begin{equation}
Y_k = \Bigg(\frac{\Psi_k(z,w)}{\big(\det\mathcal{Z}\big)^\lambda}\Bigg)\Bigg|_{\,\smallthickbar{z}\cdot w = 0}
\end{equation}
are the spherical harmonics.
\end{quote}

This can be verified directly by parameterizing the constraint $\smallthickbar{z}\cdot w = 0$ as $w^i = \varepsilon^{ij}\smallthickbar{z}_j$. In this parameterization,
\begin{equation}
Y_k = \frac{\tilde{T}_{i_1\ldots i_k}^{j_1\ldots j_k} z^{i_1}\smallthickbar{z}_{j_1}\ldots z^{i_k}\smallthickbar{z}_{j_k}}{|z|^{2k}}\,,
\end{equation}
where $\tilde{T}$, obtained from $T$ and $\varepsilon$ tensors, is a traceless tensor with respect to contraction of upper and lower indices. Such functions are the complex analogue of the classic characterization of spherical harmonics as restrictions of harmonic homogeneous polynomials to the sphere (see \cite{Bykov:2023afs} and the Theorem 1 in \cite{Bykov:2023uwb}). The underlying message is that our formalism effectively replaces a problem in \textit{harmonic analysis} with one in \textit{complex analysis} and \textit{algebraic geometry}.

\subsubsection{Monopole field on $\mathbb{S}^2$}
Consider a slight modification of the two-site XXX $\mathbf{SU}(2)$ spin chain discussed in the previous Section. Suppose everything remains as before, except that the spins at the two sites are no longer equal; i.e., the Hilbert space is $\mathsf{V}_{\frac{\lambda}{2}}\otimes\mathsf{V}_{\frac{\lambda+\mathfrak{q}}{2}}$, where $\mathfrak{q}\in\mathbb{Z}_{\geq 0}$ is an additional parameter. Equivalently, we replace the constraints \eqref{QuantumSphereConstraint} by
\begin{equation}\label{MagneticConstraints}
a^\dagger\cdot a = \lambda\mathds{1}\,,\qquad b^\dagger\cdot b= (\lambda+\mathfrak{q})\mathds{1}\,.
\end{equation}
A natural question then arises: does this modified spin chain correspond to some sigma model in the $\lambda\rightarrow\infty$ limit (with $\mathfrak{q}$ fixed)? The answer is affirmative. Repeating the calculations of the previous Section with only minor adjustments shows that, in the large‑spin limit the model coincides with the $\mathbb{CP}^1$ sigma model coupled to a magnetic monopole field of charge $\mathfrak{q}$. This means that the curvature of the magnetic connection $\mathcal{A}$ is given by $\dif\mathcal{A} = \mathfrak{q}\,\omega_{\text{FS}}$, where $\omega_{\text{FS}}$ is the (suitable normalized) Fubini–Study form on the sphere.

The spectrum of this model can again be obtained algebraically. Eigenfunctions $\Psi^{\mathfrak{q}}_k$, analogous to those in \eqref{EigenfunctionsSphere}, are constructed with the only difference that $\Psi^{\mathfrak{q}}_k$ contains $k$ quanta of type $a$ and $k+\mathfrak{q}$ quanta of type $b$. For instance, the ground state is
\begin{equation}
\Psi^\mathfrak{q}_0 = T_{i_1\ldots i_{\mathfrak{q}}}b_{i_1}^\dagger\ldots b_{i_\mathfrak{q}}^\dagger\big(\varepsilon_{ij}a^\dagger_i b^\dagger_j\big)^{\lambda}|0\rangle\quad\in\quad\mathsf{V}_{\frac{\mathfrak{q}}{2}}\,.
\end{equation}
This reproduces the Lowest Landau Level (LLL) quantization \cite{Ivanov:2007sf}, whose mathematical counterpart is known as Berezin–Toeplitz quantization \cite{Schlichenmaier:2010ui}. One readily checks that $H\Psi^{\mathfrak{q}}_k = k(k+1+\mathfrak{q})\Psi^{\mathfrak{q}}_k$, which coincides with the spectrum of the Bochner (magnetic) Laplacian on the sphere acting on sections of the line bundle $\mathscr{O}(\mathfrak{q})$ \cite{10.2748/tmj/1178228026}.

\subsection{Flag manifolds and $\mathbf{SU}(n)$ spin chains}\label{FlagSection}
In this Section we present results concerning a particle on a flag manifold and the corresponding spin chain. Recall that the flag manifold $\mathcal{F}_{n_1,\ldots, n_k}$ is the moduli space of $k$ mutually orthogonal linear subspaces in $\mathbb{C}^n$ of complex dimensions $n_i$, where $n = n_1+\ldots+n_k$ \cite{Affleck:2021vzo}. To simplify the discussion, we will primarily focus on the case of complete flags $\mathcal{F}_n$, where each $n_i = 1$.

The required Lagrangian embedding\footnote{For general flags $\mathcal{F}_{n_1,\ldots, n_k}$ one must consider an embedding into a product of Grassmannian manifolds $\mathrm{Gr}(n_1,n)\times\ldots\times\mathrm{Gr}(n_k,n)$, where every $\mathrm{Gr}(n_i,n)$ equipped with the unique (up to a scale) $\mathbf{SU}(n)$ invariant natural symplectic form which generalizes the Fubini-Study form. The geometric quantization of the corresponding $\mathbf{SU}(n)$ orbits yields $\mathfrak{su}(n)$ representations labeled by rectangular Young diagrams.} takes the form
\begin{equation}\label{FlagLagrEmb}
\mathcal{F}_n\hookrightarrow\Big(\underbrace{\mathbb{CP}^{n-1}\times\ldots\times\mathbb{CP}^{n-1}}_{n~\text{times}}, \omega_\lambda\Big)\,,\quad \omega_\lambda = i\lambda\big(\dif\smallthickbar{z}_1\wedge\dif z_1+\ldots+\dif\smallthickbar{z}_n\wedge\dif z_n\big)\,,
\end{equation}
where each $z_i\in\mathbb{C}^n$ with the identification $z_i\sim e^{i\varphi_i}z_i$ and the normalization $|z_i|^2 = 1$ defines the $i$-th copy of $\mathbb{CP}^{n-1}$. The embedding is constructed as follows. Since a point in $\mathbb{CP}^{n-1}$ corresponds to a one‑dimensional subspace of $\mathbb{C}^n$, a point in $\big(\mathbb{CP}^{n-1}\big)^{\times n}$ is an ordered set of $n$ such subspaces. The complete flag manifold is embedded by imposing the condition that these subspaces are mutually orthogonal. Moreover, this embedding is Lagrangian with respect to the sum of Fubini–Study forms with equal normalizations. This leads to the following generalization of \eqref{TheorCotB}:

\begin{quote}
\underline{\textbf{Theorem.}} (\cite{Bykov:2024tvb}) There exists a symplectomorphism
\begin{equation}\label{FlagCotB}
\mathrm{T}^\ast\mathcal{F}_n \simeq\lim_{\lambda\rightarrow\infty}\frac{\big(\mathbb{CP}^{n-1}\times\ldots\times\mathbb{CP}^{n-1},\omega_\lambda\big)}{\{\det\mathcal{Z}=0\}}\,,
\end{equation}
where $\mathcal{Z}$ is the $n\times n$ matrix whose columns are the vectors $z_i$.
\end{quote}

The proof of this theorem essentially repeats the computation given in Section \ref{SectionSphere}.

We now quantize both sides of \eqref{FlagCotB}. The complex projective spaces $\mathbb{CP}^{n-1}$ are particular coadjoint orbits of $\mathbf{SU}(n)$; their geometric quantization yields the space of holomorphic sections of the line bundle $\mathscr{O}(\lambda)$, which carries the structure of the fully symmetric $\mathfrak{su}(n)$ representation. Consequently, the quantum counterpart of \eqref{FlagCotB} reads
\begin{equation}
L^2(\mathcal{F}_n)\simeq\lim_{\lambda\rightarrow\infty}\Big(\underbrace{\mathsf{V}_{\frac{\lambda}{2}}\otimes\ldots\otimes\mathsf{V}_{\frac{\lambda}{2}}}_{n~\text{times}}\Big)\,,
\end{equation}
where $\mathsf{V}_{\frac{\lambda}{2}}$ denotes the fully symmetric $\mathfrak{su}(n)$ representation corresponding to a Young diagram consisting of a single row of length $\lambda$.

The simplest classical Hamiltonian that attains a global minimum on the Lagrangian image of $\mathcal{F}_n$ is
\begin{equation}\label{SUnHamiltonian}
H = \sum\limits_{i>j = 1}^n \alpha_{ij}|\smallthickbar{z}_i\cdot z_j|^2\,,\qquad\alpha_{ij}\in\mathbb{R}\,.
\end{equation}
It can also be shown that the classical spin chain defined by the right‑hand side of \eqref{FlagCotB} with the Hamiltonian \eqref{SUnHamiltonian} is equivalent to a classical particle on $\mathcal{F}_n$ moving in the metric
\begin{equation}\label{FlagMetric}
\dif s^2 = \sum\limits_{i> j = 1}^n \frac{|\smallthickbar{u}_i\cdot \dif u_j|^2}{\alpha_{ij}}\,,\qquad \smallthickbar{u}_i\cdot u_j = \delta_{ij}\,,\qquad u_i\sim e^{i\phi_i} u_i\,,
\end{equation}
where $u_i$ defines the $i$'th ordered line in the complete flag manifold. This form of the metric also requires that the coefficients $\alpha_{ij}$ be strictly positive.

Using the Schwinger–Wigner oscillator representation discussed in Section \ref{SectionSphere}, the quantization of \eqref{SUnHamiltonian} yields the spin‑$\frac{\lambda}{2}$ $\mathbf{SU}(n)$ Hamiltonian
\begin{equation}\label{XXXSUnHamiltonian}
H = \frac{1}{2}\sum\limits_{i<j = 1}^n \alpha_{ij}\sum\limits_{k = 1}^{\mathsf{dim}\,\mathfrak{su}(n)}S^k_iS^k_j\,,
\end{equation}
where $S^k_i$ is the $k$'th generator of $\mathfrak{su}(n)$ in the spin‑$\frac{\lambda}{2}$ representation\footnote{If different symmetric representations are taken on each site, one obtains the flag‑manifold sigma model with up to $n-1$ independent monopole charges, generalizing the case of Section \ref{SectionSphere}. The number of monopoles can be explained by the fact that $H^2\big(\mathcal{F}_n\big)\simeq\mathbb{Z}^{n-1}$, where the generators are pullbacks of 2‑forms proportional to the Fubini–Study form on the projective spaces under the natural embedding $\mathcal{F}_n\hookrightarrow\big(\mathbb{CP}^{n-1}\big)^n$. Note that one has $\mathbb{Z}^{n-1}$ rather than $\mathbb{Z}^{n}$ because the Lagrangian embedding \eqref{FlagLagrEmb} imposes one relation $\omega_\lambda = 0$ among the cohomology generators.} at the $i$'th site of the spin chain. Note that this yields a highly non‑local, ``all‑to‑all'' spin chain\footnote{Such models are sometimes referred to as ``spin nets''.}. According to the general logic outlined in Section \eqref{GeneralQuantSection}, the spectrum of the Hamiltonian \eqref{XXXSUnHamiltonian} should reproduce the spectrum of the Laplace–Beltrami operator in the metric \eqref{FlagMetric} in the limit $\lambda\rightarrow\infty$. In fact, this reformulation of the spectral problem is highly productive. For example, using the Bethe Ansatz method, the spectrum of $-\triangle$ on $\mathcal{F}_3$ was recently obtained for a general invariant metric \cite{BykovGaudin}, whereas previously only the simplest special case $\alpha_{12} = \alpha_{23} = \alpha_{13}$ was known.

\subsection{Particle on hyperbolic plane and $\mathbf{SL}(2,\mathbb{R})$ spin chain}\label{HyperbolicSection}
In this brief section we present another example, which involves coadjoint orbits of the non‑compact group $\mathbf{SU}(1,1)\simeq\mathbf{SL}(2,\mathbb{R})$. For simplicity we treat only the kinematical aspect of the construction. Even this limited discussion, however, provides a simple way to recover quite non‑trivial results about infinite‑dimensional representations.

The Poincaré disk model of the hyperbolic (Lobachevsky) plane $\mathbb{H}$ is the unit disc $\mathbb{D}\subset\mathbb{C}$ equipped with the metric
\begin{equation}
\dif s^2 = \frac{\dif {q}\dif {\smallthickbar{q}}}{\big(1-|q|^2\big)^2}\,,\qquad |q|<1\,.
\end{equation}
Here $q$ is the hyperbolic analogue of an inhomogeneous coordinate. There are also two “homogeneous” parameterizations related to the coset representation $\mathbb{H}\simeq\frac{\mathbf{SU}(1,1)}{\mathbf{U}(1)}$. The first, which we denote by $\mathbb{H}_+$, is defined by a complex two‑component vector $z\in\mathbb{C}^2$ identified up to a phase, $z\sim e^{i\varphi}z$, and satisfying the indefinite normalization $\smallthickbar{z}\eta z = 1$, where $\eta = \mathrm{diag}[1,-1]$ is the flat Minkowski metric. The second, $\mathbb{H}_-$, is given by a two‑component vector $w$ with $w\sim e^{i\phi}w$ and $\smallthickbar{w}\eta w = -1$. As smooth manifolds, $\mathbb{H}_+$ and $\mathbb{H}_-$ are clearly symplectomorphic, but the $\mathbf{SU}(1,1)$ actions -- which rotate $z$ and $w$ simply by matrix multiplication -- are not equivalent\footnote{In other words, the symplectomorphism is not equivariant.}, which affects quantization.

The hyperbolic plane is itself a coadjoint orbit of $\mathbf{SU}(1,1)$. Consider therefore the classical spin‑chain phase space $\big(\mathbb{H}_+\times\mathbb{H}_-,\omega_\lambda\big)$, where the symplectic form is
\begin{equation}
\omega_\lambda = i\lambda\big(\dif{\smallthickbar{z}}\wedge\eta\dif z-\dif{\smallthickbar{w}}\wedge\eta\dif w\big)\,.
\end{equation}
The claim is that there exists a Lagrangian embedding
\begin{equation}
\mathbb{H}\hookrightarrow\big(\mathbb{H}_+\times\mathbb{H}_-,\omega_\lambda\big)\qquad\text{defined by}\qquad\smallthickbar{z}\eta w = 0\,.
\end{equation}
Using methods analogous to those of Section \ref{SectionSphere}, one can prove that this embedding induces the symplectomorphism\footnote{The key technical tool required to implement the appropriate change of variables is the indefinite polar decomposition \cite{Bolshakov}.}
\begin{equation}\label{HyperbolicSympl}
\mathrm{T}^\ast\mathbb{H}\simeq \big(\mathbb{H}_+\times\mathbb{H}_-,\omega_\lambda\big)\,.
\end{equation}
This is an example where neither the large $\lambda$ limit nor the excision of any submanifold is required. The reason is as follows. As discussed in Section \ref{SectionSphere}, the large $\lambda$ limit was needed to obtain a “symplectomorphism” between compact coadjoint orbits and the non‑compact cotangent bundle, and cutting out the determinantal variety was necessary to trivialize certain quantum line bundles. Here, neither step is needed because the orbits themselves are non‑compact and contractible.

Quantizing the relation \eqref{HyperbolicSympl} leads to an interesting representation‑theoretic result:
\begin{equation}
L^2(\mathbb{H})\simeq\mathcal{D}_s^+\otimes\mathcal{D}_s^-\,,
\end{equation}
where $\mathcal{D}_s^\pm$ denote the positive and negative discrete‑series representations \cite{KitaevSL2R} of the same (but otherwise arbitrary admissible) “spin” $s = \frac{1}{2}(1+\lambda)$. This reproduces a not‑so‑old result of \cite{Repka}. The dynamical aspects of this correspondence are left for future work~\cite{JointPaper}.

\section*{Acknowledgments}
I am immensely grateful to Dmitri Bykov and Andrew Kuzovchikov; without their essential discussions and contributions, this work would never have been possible. I also wish to thank Mikhail Markov, Ivan Sechin and Nikolai Tyurin for valuable discussions. This work originated from notes prepared for presentations at the St.~Petersburg Department of the Steklov Mathematical Institute (PDMI) and the Beijing Institute of Mathematical Sciences and Applications (BIMSA), and I thank Anton Selemenchuk and Andrii Liashyk for making these talks possible. Part of this work was carried out during a visit to BIMSA, and I am grateful to Anton Pribytok for his invitation. The work is supported by Russian Science Foundation grant № 25-72-10177.

\printbibliography
\end{document}